\documentclass[preprint]{aastex} 

\shorttitle{Long Term Evolution} 
\shortauthors{F. C. Adams} 
 
\newcommand{\be}{\begin{equation}}
\newcommand{\ee}{\end{equation}}
\newcommand{\ebar}{{\langle e \rangle}}

\newcommand{\taucirc}{{\tau_{cir}}} 
\newcommand{\taups}{{ \tau_{P\ast}}}

\newcommand\tz{ {\hat t} } 
 
\def\lta{\,\raise 0.3 ex\hbox{$ < $}\kern -0.75 em
 \lower 0.7 ex\hbox{$\sim$}\,}
\def\gta{\,\raise 0.3 ex\hbox{$ > $}\kern -0.75 em
 \lower 0.7 ex\hbox{$\sim$}\,} 

\begin{document}
 
\title{Long Term Evolution of Close Planets \\ 
Including the Effects of Secular Interactions} 
 
\author{Fred C. Adams$^{1,2}$ and Gregory Laughlin$^3$} 

\affil{$^1$Michigan Center for Theoretical Physics \\
Physics Department, University of Michigan, Ann Arbor, MI 48109} 
 
\affil{$^2$Astronomy Department, University of Michigan, Ann Arbor, MI 48109}

\affil{$^3$Lick Observatory, University of California, Santa Cruz, CA 95064} 
 
\begin{abstract} 

This paper studies the long term evolution of planetary systems
containing short-period planets, including the effects of tidal
circularization, secular excitation of eccentricity by companion
planets, and stellar damping. For planetary systems subject to all of
these effects, analytic solutions (or approximations) are presented
for the time evolution of the semi-major axes and eccentricities.
Secular interactions enhance the inward migration and accretion of hot
Jupiters, while general relativity tends to act in opposition by
reducing the effectiveness of the secular perturbations. The analytic
solutions presented herein allow us to understand these effects over a
wide range of parameter space and to isolate the effects of general
relativity in these planetary systems.

\end{abstract}

\keywords{Stars: Planetary systems} 

\section{Introduction} 

Starting with the discovery of extrasolar planets (Mayor \& Queloz
1995; Marcy \& Butler 1996), a substantial fraction of the planetary
orbits have been found close to their stars, with periods $P \sim 4$
days. These objects are often referred to as ``hot Jupiters''.  With
$\sim200$ planets detected to
date\footnote{http://www.ucolick.org/\~{$\,$}laugh/}, the distribution of
orbital periods shows a measurable pile-up at periods $P$ = 3--5 days,
i.e., roughly 10 percent of the currently detected planets have
periods $P < 5$ days. Although the close planets are the most easily
detected, this finding is not a selection effect: The current
statistics indicate that 1.2 percent of all FGK stars have hot
Jupiters within 0.1 AU of their stars (Marcy et al. 2005).  These hot
Jupiters are subject to tidal interactions with their central stars
(e.g., Goldreich \& Soter 1963), and this star-planet coupling can
influence the long term evolution of these systems. If the system
contains additional bodies, then planet-planet interactions can also
affect the long term evolution of the close planets.  In our companion
paper (Adams \& Laughlin 2006a; hereafter Paper I), we present a
treatment of secular interactions for (non-resonant) systems with
multiple planets.  In this paper, we combine this formulation of
secular interactions with star-planet interactions to study the long
term fate and evolution of close planets.

The basic theory of secular interactions, as applied to extrasolar
planetary systems and used herein, is reviewed in Paper I (see also
Murray \& Dermott 1999; hereafter MD99). These interactions allow for
planetary orbits to exchange angular momentum so that orbital
eccentricities change on secular time scales that are long compared to
both the orbit time and observational monitoring times (tens of
years), but short compared to the ages of the systems (a few Gyr).
The characteristic secular interaction time scales provide a simple
metric of the importance of planet-planet perturbations within a given
system.  For a collection of observed multi-planet systems, these
secular time scales fall in the range 100 -- 50,000 yr, as listed in
Table 1 of Paper I.  On longer time scales, tidal interactions with
the star act to circularize close orbits and can lead to continued
inward migration.  This process might have a bearing on the observed
pile-up of planets with periods of 3--4 days and can lead to
considerable energy input into the planetary atmospheres.  Although
this process has been considered previously (e.g., Trilling 2000;
Bodenheimer et al. 2001, 2003; Yu \& Goldreich 2002; Mardling \& Lin
2002, 2004; Gu et al. 2003; Faber et al. 2005), this paper takes the
additional step of providing analytic expressions for the evolution
time for various classes of systems, including both tidal damping and
secular interactions.  This work thus provides additional analytic
insight into the problem as well as elucidating the role played by
secular interactions.  The version of secular interaction theory
formulated in Paper I includes the leading order corrections for
general relativity (GR), which causes the periasta of planets to
precess forward in their orbits.  As a result, this semi-analytical
treatment also allows us to explicitly delineate the role played by
general relativity in the long term evolution of close planets.

For a single planet in a close orbit, the tidal circularization effect
is usually written in terms of the time scale for eccentricity decay,
$\taucirc = -e / {\dot e}$. For close orbits in multiple planet
systems, however, the eccentricity is excited via interactions that we
model according to secular theory. The first approximation to the long
term behavior is to assume that the orbit decays with constant angular
momentum and that the semi-major axis decreases on the same time
scale, i.e.,
\be
{ {\dot a} \over a} = {2 e {\dot e} \over 1 - e^2} = - 
{2 e^2 \over (1 - e^2)}\, {1 \over \taucirc} \, , 
\ee 
where the time scale $\taucirc$ for circularization 
can be written in the form 
\be
\taucirc \approx {4 Q_P \over 63} \Bigl( {a^3 \over GM_\ast} 
\Bigr)^{1/2} {m_P \over M_\ast} \Bigl( {a \over R_P} \Bigr)^5 \, 
(1 - e^2)^{13/2} [ F(e^2) ]^{-1} \, ,  
\label{eq:circone} 
\ee  
where $Q_P \approx 10^5 - 10^6$ is the tidal quality factor and $R_P$
is the planet radius. For extrasolar planets, the quality factor $Q_P$
and the radius $R_P$ depend sensitively on the planetary mass,
temperature, and composition (Bodenheimer et al. 2003).  This general
form for the time circularization scale is well known (e.g., Goldreich
\& Soter 1966), but includes additional factors to account for the
effects of nonzero eccentricity (see Hut 1981). The choice $F \approx
1 + 6 e^2 + {\cal O}(e^4)$ provides a reasonable approximation to the
results for close binaries (Hut 1981) for moderate eccentricities.
Inserting representative values, and taking the limit $e \to 0$, we
write the circularization time scale in the form
\be
\tau_0 = \taucirc (e=0) \approx  \, \, 
1.6 \, {\rm Gyr} \, \, \Bigl( {Q_P \over 10^6} \Bigr) 
\, \, \Bigl( {m_P \over m_J} \Bigr) \, \, 
\Bigl( {M_\ast \over M_\odot} \Bigr)^{-3/2}  
\, \, \Bigl( {R_P \over R_J} \Bigr)^{-5} 
\, \, \Bigl( {a \over 0.05 \, {\rm AU}} \Bigr)^{13/2} 
\, \, . 
\label{eq:circtwo} 
\ee 
Throughout this paper we work in terms of the dimensionless time scale  
\be 
\tz \equiv { t / \tau_0} \, . 
\ee

These systems are also affected by a stellar damping effect in which
energy is dissipated in the star due to tides raised by the
planet. The effectiveness of this process is determined by an
analogous parameter $Q_\ast$, the tidal quality factor of the star.
The net result of this process is to cause the semi-major axis of the
inner planet to decay on a time scale $\taups$ (again, see Goldreich
\& Soter 1966, Hut 1981) that can be written in the form
\be
\taups = \tau_0 \Gamma^{-1} \Bigl( {a \over a_0} \Bigr)^{13/2} \, , \qquad 
{\rm where} \qquad \Gamma \equiv {2\over 7} \, 
\Bigl( {R_\ast \over R_P} \Bigr)^5 \, 
\Bigl( {Q_P \over Q_\ast} \Bigr) \, 
\Bigl( {m_P \over M_\ast} \Bigr)^2 \, . 
\label{eq:taupdef} 
\ee 
As defined here, $\Gamma$ is generally a small parameter. For
Jupiter-like planets, $R_P \sim 0.1 R_\ast$ and $m_P \sim 10^{-3}
M_\ast$ so that $\Gamma_J \sim 0.03 (Q_P/Q_\ast)$. For Neptune-like
planets, we find $\Gamma_{Nep} \sim 0.015 (Q_P/Q_\ast)$, but we expect
the $Q_P$ value to be much smaller so that $\Gamma_{Nep} \ll \Gamma_J$. 

This paper considers hot Jupiters subject to tidal circularization
effects (with strength determined by $Q_P$), both in single planet
systems (\S 2.1) and in two-planet systems where the eccentricity of
the inner planet is excited through secular interactions (\S 2.2).
Next we consider hot Jupiter systems with additional stellar damping
effects (with strength determined by $Q_\ast$), both for single planet
systems (\S 3.1) and two-planet systems (\S 3.2). Our results are
summarized in \S 4 along with a discussion of their ramifications.

\section{Long Term Evolution with No Stellar Damping} 

The basic goal of this section is to provide an analytic understanding
of the long term behavior of the hot Jupiter systems in the absence of
stellar damping terms (note that stellar damping is included in the
following section).  For single planet systems that experience tidal
forces only, we define $g(t) \equiv a_1(t)/a_1(0)$. Similarly, for two
planet systems, we define $f(t) \equiv a_1(t)/a_1(0)$, where the inner
planet experiences circularization from the central star and
eccentricity excitation from the other planet.

\subsection{One planet system with no stellar damping} 

For a one planet system, the evolution is described by two coupled
equations of motion. If we work to the same order of approximation 
in $e^2$, these equations take the form  
\be 
{dg \over d\tz} = - 2 e^2 g^{-11/2} \qquad {\rm and} \qquad 
{de \over d\tz} = - e g^{-13/2} \, , 
\label{eq:start} 
\ee
which be be combined to form the second order differential equation 
\be 
\bigl( {dg \over d\tz} \bigr)^{-1} {d^2 g \over d\tz^2} + 
{11 \over 2g} {dg \over d\tz} = - 2 g^{-13/2} \, . 
\ee 
The first integral of this equation can also be found, i.e., 
\be 
{dg \over d\tz} = - 2 g^{-11/2} \bigl( e_{1(0)}^2 + \ln g \bigr) \, ,
\ee 
where $e_{1(0)}$ is the eccentricity of the inner planet at $t=0$. In
contrast to the two planet case, where continued eccentricity forcing
by the second planet leads to continued evolution, this system
asymptotically approaches a minimum value of semi-major axis, i.e., a
minimum value of $g$ given by $g_\infty = \exp [ - e_{1(0)}^2 ]$. 
Combining the two expressions in equation (\ref{eq:start}), we can 
also solve directly for the eccentricity as a function of the
factor $g$, i.e., $e_1^2 = e_{1(0)}^2 + \ln g$.  This expression does
not satisfy conservation of angular momentum exactly because of the
approximation made at the start (where we work to only leading order
in $e^2$).

If we include the additional factors to enforce conservation of 
angular momentum, the equations of motion take the form 
\be 
{dg \over d\tz} = - 2 {e^2 \over 1 - e^2} g^{-11/2} 
\qquad {\rm and} \qquad 
{de \over d\tz} = - e g^{-13/2} \, , 
\ee
and the resulting second order differential equation becomes 
\be 
\bigl( {dg \over d\tz} \bigr)^{-1} {d^2 g \over d\tz^2} + 
{11 \over 2g} {dg \over d\tz} = - 2 g^{-13/2} {g \over 1 - e_0^2} \, .
\ee 
Here we have eliminated the eccentricity dependence using conservation
of angular momentum, i.e., $g (1 - e^2) = (1 - e_0^2)$ = $J^2$ = {\sl
constant}.  This second order differential equation can be integrated
once to find the time as a function of $g$, i.e., 
\be 
\tz = {J^2 \over 2} \int_g^1 {g^{11/2} dg \over g - J^2} \, . 
\label{eq:jint} 
\ee 
The integral can also be evaluated to find the implicit solution 
$$ 
\tz = {J^2 \over 11} \bigl(1 - g^{11/2} \bigr) 
+ {J^4 \over 9} \bigl(1 - g^{9/2} \bigr) 
+ {J^6 \over 7} \bigl(1 - g^{7/2} \bigr) 
+ {J^8 \over 5} \bigl(1 - g^{5/2} \bigr) + 
$$
\be
{J^{10} \over 3} \bigl(1 - g^{3/2} \bigr) + 
J^{12} \bigl(1 - g^{1/2} \bigr) + {1 \over 2} 
J^{13} \ln \Big| {1 - J \over 1 + J} \cdot 
{g^{1/2} + J \over g^{1/2} - J } \Big| \, . 
\label{eq:jsol} 
\ee
Both the time integral (eq. [\ref{eq:jint}]) and the solution
(eq. [\ref{eq:jsol}]) indicate that the planet only reaches its 
final location (given by $g=J^2$) asymptotically in time 
($t \to \infty$). 

\subsection{Two planet system with no stellar damping} 

For the long term evolution of a two planet system, the eccentricity
of the inner planet is driven by the interactions with the outer
planet. Within this set of approximations, the evolution is given by
the time-averaged equation 
\be 
{ {\dot f} \over f} = - 2 \langle e^2 \rangle 
\, \langle {1 \over \taucirc} \rangle \, = 
- 2 \langle e^2 \rangle \, f^{-13/2} \, , 
\ee 
where the angular brackets denote time averages over an intermediate
time scale that is long compared to the secular time scales (e.g., see
Table 1 of Paper I) and short compared to the circularization time
scales (eq. [\ref{eq:circtwo}]). As the system evolves, the semi-major
axis of the inner planet decreases, which in turn causes the secular
averaged square eccentricity $\langle e^2 \rangle$ to decrease. 
Notice that by time-averaging the square of the eccentricity and the
circularization time scale as separate quantities, we are making an
approximation that limits the accuracy of this treatment to ${\cal
O}(e^2)$. In practice, however, we evaluate $\taucirc$ in the $e \to
0$ limit, and the secular interaction theory used here is only
accurate to second order in $e$ (Paper I; MD99), so that this order of
approximation is consistent with our general framework.

In the absence of other effects, the equation of motion for 
the two planet system can be written in the form  
\be 
{df \over d\tz} = - 4 \eta^2 f^{-11/2} \, , 
\ee
where the eccentricity excitation amplitude $\eta$ takes the 
approximate form 
\be
\eta^2 \approx {25 \over 16} { e_{2(0)}^2 \alpha_0^2 f^2 \over 
(1 + \Pi_0 f^{-3} - \delta \sqrt{\alpha_0 f} )^2 
+ (25 / 4) \delta \alpha_0^{5/2} f^{5/2} } \, , 
\ee 
where we have defined $\delta \equiv m_1 / m_2$ and $\alpha_0 \equiv
a_1(0) / a_2 (0)$. The dimensionless parameter $\Pi_0$ = $4 G M_\ast^2
a_2^3 / (m_2 c^2 a_1^4)$ provides a measure of the importance of
general relativity in secular interactions, where the semi-major axes
are evaluated at $t=0$ (from eq. [17] of Paper I; see also Adams \&
Laughlin 2006b).  The quantity $e_{2(0)}$ is the eccentricity of the
second planet evaluated at $t=0$ (although this eccentricity is
assumed constant in this set of approximations). It is useful to
collect the constants into the dimensionless composites
\be
A \equiv {25 \over 4} e_{2(0)}^2 \alpha_0^2 \, , \qquad 
B \equiv \delta \sqrt{\alpha_0} \, , \qquad {\rm and} \qquad 
C \equiv {25 \over 4} \delta \alpha_0^{5/2} \, . 
\ee 
With these definitions, the equation of motion can be written in the form 
\be
{df \over d\tz} = - {A f^{5/2} \over (\Pi_0 + f^3 - B f^{7/2})^2 
+ C f^{17/2} } \, . 
\label{eq:dfdt} 
\ee 
By absorbing the leading coefficient $A$ into the time variable, we can
define the relevant time scale $\tau_S$ for the evolution of this
system, i.e., 
\be
\tau_S \equiv {4 \tau_0 \over 25 \alpha_0^2 e_{2(0)}^2 } \, 
= {\tau_0 \over A} \, . 
\ee 
This time scale is often much longer than the circularization time
$\tau_0$ that applies for single planet systems. For example, using
the observed parameters for the inner two planets of the Ups And
system, we find $A \approx 0.0022$ so that $\tau_S/ \tau_0 \approx 454$. 

The equation of motion (\ref{eq:dfdt}) can be integrated to provide
the solution in implicit form
$$
A \tz = {2 \over 9} (1 - f^{9/2}) + {4 \over 3} \Pi_0 (1 - f^{3/2}) 
+ {2 \over 3} \Pi_0^2 (f^{-3/2} - 1) + {C \over 7} (1 - f^7) 
$$
\be 
+ {2 \over 11} B^2 (1 - f^{11/2}) - \Pi_0 B (1 - f^2) - 
{2 \over 5} B (1 - f^5) \, . 
\label{eq:twosol} 
\ee 
In the limit $B,C \ll 1$ (which often holds), this solution 
can be simplified further to the form: 
\be
A \tz = {2 \over 9} (1 - f^{9/2}) + {4 \over 3} \Pi_0 (1 - f^{3/2}) 
+ {2 \over 3} \Pi_0^2 (f^{-3/2} - 1) \, . 
\label{eq:twosolR} 
\ee 
From this expression, we can read off the asymptotic behavior. As the
time $\tz$ increases, $f \ll 1$, and the third term in the equation
dominates. As a result, as $\tz \to \infty$, the function $f(\tz)$
approaches the limiting form 
\be
f (\tz) \, \to \, \Pi_0^{4/3} \, (3A\tz/2)^{-2/3} \, . 
\ee
Unlike the case of a single planet system (see \S 2.1), the migrating 
planet can continue to lose energy and can become arbitrarily close 
to the origin. In practice, however, once the planet reaches the 
stellar surface where $f = f_{min} = R_\ast/a_1(0)$, the planet will 
be destroyed and evolution will be over. The time required for the 
planet to reach the stellar surface is thus given by 
\be
A \tz_\ast = {2 \over 9} (1 - f_{min}^{9/2}) + {4 \over 3} \Pi_0 
(1 - f_{min}^{3/2}) + {2 \over 3} \Pi_0^2 (f_{min}^{-3/2} - 1) \, . 
\label{eq:fulltime} 
\ee 
Notice the important role played by general relativity in this
setting. In the absence of relativistic corrections, the (leading
order) solution would have the form $f(\tz) = (1 - 9A\tz/2)^{2/9}$,
which reaches $f=0$ in the relatively short time $\tz_\ast = 2/9A$.
Relativistic precession thus acts to keep the planet from being
accreted by the star. This claim can be quantified by inserting
typical values of interest; let $a_1(0) = 0.05$ AU and $R_\ast$ =
1.0 $R_\odot$ so that the total evolution time given by equation
(\ref{eq:fulltime}) can be written $9A\tz_\ast/2 \approx 1 + 6
\Pi_0 + 105 \Pi_0^2$.  With these values, e.g., the condition for
relativistic effects to dominate becomes $\Pi_0 \gta 0.073$.

The portion of parameter space for which two planet systems lead to
significant (short) accretion times is depicted in Figure
\ref{fig:longterm1}.  In this application, we assume that the inner
planet is a hot Jupiter, with mass $m_1 = m_J$, radius $R_P = R_J$,
and starting semi-major axis $a_1$ = 0.05 AU. The eccentricity of the
inner planet cycles through $e_1$ = 0, although $\ebar \ne 0$ due to
secular interactions with the second planet. The stellar mass $M_\ast$
= 1.0 $M_\odot$ and the tidal quality factor $Q_P = 10^6$ so that the
circularization time scale $\tau_0$ = 1.6 Gyr (see eq.
[\ref{eq:circtwo}]).  The figure shows the orbital elements of the
second planet $(a_2,e_2)$ required to drive the hot Jupiter into the
stellar photosphere over a fiducial time scale of 5 Gyr. The accretion
time depends on the mass of the second planet, which is assumed to
vary over the range $m_2 = 1 - 5 m_J$, corresponding to the five solid
curves shown in Figure \ref{fig:longterm1}.  The region of the $a-e$
plane above the curves represents solar systems in which the inner
planet would be driven into the star on time scales less than 5 Gyr.
The dashed curve delineates the (much larger) region of parameter
space for which the inner planet would be driven into the star in the
absence of general relativity (where $m_2 = m_J$). Comparison of the
dashed curve with the upper solid curve thus illustrates how GR acts
to prevent the accretion of planets.

\section{Long Term Evolution with Stellar Damping Term} 

Given the solutions discussed above, the next correction is to
consider the dissipation of energy that occurs due to the planet
raising tides on the star. This section presents solutions for the
long term evolution of these systems, including both tidal
circularization (due to energy dissipated in the planet via $Q_P$) and
orbital damping (due to energy dissipated in the star via $Q_\ast$).
We present solutions for single planets systems (\S 3.1) and for two
planet systems in which the bodies are also subject to secular
interactions (\S 3.2).

\subsection{One planet system with stellar damping term} 

In the presence of stellar damping, the equations of motion for 
the one planet system take the form 
\be 
{dg \over d\tz} = - 2 {e^2 \over 1 - e^2} g^{-11/2} - \Gamma g^{-11/2} 
\qquad {\rm and} \qquad 
{de \over d\tz} = - e g^{-13/2} \, . 
\ee
In the limit of zero eccentricity, the stellar damping term acts to 
decrease the semi-major axis of the planet according to $g(\tz)$ =
$(1 - 13 \Gamma \tz / 2)^{2/13}$; the system thus has a ``natural'' 
damping time scale of $\tz_{\rm c} = 2/13\Gamma$. Combining the 
above equations, we can solve directly for $g(e)$, 
\be 
g = \Bigl( {e \over e_0} \Bigr)^\Gamma 
\Bigl( {1 - e_0^2 \over 1 - e^2} \Bigr) \, . 
\label{eq:gofe} 
\ee 
We can insert this form back into the eccentricity evolution 
equation and solve for the time as a function of $e$, 
\be 
\tz = (1 - e_0^2)^{13/2} \, e_0^{-13\Gamma/2} \, 
\int_e^{e_0} {de \over e} \,  {e^{13\Gamma/2} \over 
(1 - e^2)^{13/2} } \, . 
\label{eq:timeint} 
\ee 
In the limit of small eccentricity, we can evaluate the integral 
to find the approximate solution 
\be 
\tz \approx (1 - e_0^2)^{13/2} \, {2 \over 13 \Gamma} \, 
\Bigl[ 1 - (e/e_0)^{13 \Gamma/2} \Bigr] \, . 
\label{eq:timesol} 
\ee 
Alternatively, we can find the approximate solution for $e(t)$, 
\be
e(t) \approx e_0 \Bigl[ 1 - {13 \over 2} \Gamma 
(1 - e_0^2)^{-13/2} \tz \Bigr]^{2/13\Gamma} \, . 
\ee
This function, in conjunction with the solution for $g(e)$ 
found above, thus specifies the semi-major axis of the planet 
as a function of time. 

The total evolution time is given by the integral of equation 
(\ref{eq:timeint}) in the limit that $e \to 0$. This time scale 
can be written in terms of the series 
\be
\tz_T = (1 - e_0^2)^{13/2} \, \sum_{n=0}^\infty 
{b_n \over n!} {e_0^{2n} \over 2n + 13 \Gamma/2} \, , 
\label{eq:sum} 
\ee 
where the coefficients $b_n$ are defined through the recursion 
relation
\be
b_{n+1} = \big( 13 \Gamma / 2 + n \big) \, b_n  
\qquad {\rm with} \qquad b_0 = 1 \, . 
\ee

These analytic results tell us a lot about the long term behavior of
the system. Equation (\ref{eq:gofe}) implies that the planet does not
reach the origin ($g \to 0$) until its eccentricity vanishes. Further,
the integral in equation (\ref{eq:timeint}) is finite for all $\Gamma
> 0$, so the evolution takes place over a finite time (equivalently,
the series in equation [\ref{eq:sum}] converges).  Finally, equation
(\ref{eq:sum}) provides a good working approximation to the total
evolutionary time scale, i.e.,
\be 
\tz_T \approx \, J^{13} \, \Bigl[ { 2 \over 13 \Gamma} +  
{13 \Gamma e_0^2 \over 4 + 13 \Gamma} + 
{13 \Gamma (2 + 13 \Gamma) e_0^4 \over 4(8 + 13 \Gamma)} \Bigr] \, . 
\label{eq:totalt} 
\ee
For many applications, the first term provides an adequate
approximation. Notice that the total evolution time is shorter than
the characteristic time scale $\taups = \tau_0/\Gamma$ (i.e., $\tz =
1/\Gamma$), as defined by equation (\ref{eq:taupdef}).  When the
leading order term is a valid approximation, the evolution time is
shorter than the characteristic time $\tz_{\rm c}$ = $2/13\Gamma$ by 
an additional factor of $J^{13} = (1 - e_0^2)^{13/2}$. For example, 
if $e_0$ = 0.28 (the median of the observed sample of extrasolar planets), 
the eccentricity causes the evolutionary time to be shorter by a factor of 
$\sim2$ compared to evolution with no initial eccentricity. 

The observed sample of extrasolar planets shows a population of
planets with periods of 3 -- 4 days, but a deficit of planets with
shorter periods. One could, in principle, explain this signature if
the shorter period planets were all accreted by their central stars.
If this explanation were true, then the total evolution time defined
above must be comparable to the stellar ages $\tau_\ast$ (which are
typically several Gyr). The requirement that all planets with
(initial) semi-major axis less than $a_0$ are accreted within the
stellar age $\tau_\ast$ can be evaluated by using the leading order
term for the evolution time (in eq. [\ref{eq:totalt}]) and the
definitions of $\tau_0$ and $\Gamma$ (eqs. [\ref{eq:circone}] and
[\ref{eq:taupdef}]). The result takes the form
\be 
\tau_\ast \ge 8.6 \, {\rm Gyr} \, \, J^{13} 
\Bigl( {m_P \over m_J } \Bigr)^{-1} 
\Bigl( {Q_\ast \over 10^6 } \Bigr)  
\Bigl( {a_0 \over 0.05 {\rm AU} } \Bigr)^{13/2} \, , 
\label{eq:tevaluate}  
\ee 
where we have assumed solar properties for the star ($M_\ast = 1
M_\odot$ and $R_\ast = 1 R_\odot$). Thus, planets of roughly Jovian
mass can be accreted in a typical stellar age (5 Gyr) provided that
$Q_\ast \sim 10^6$ (which is a reasonable value, implied by
considerations of eccentricity damping in close binaries, e.g., Hut
1981). Because of the sensitive dependence on semi-major axis $a$,
planets that are accreted spend relatively little time with shorter
periods, and, the predicted cutoff in observed period is relatively
sharp.  Note that $\tau_\ast \propto a_0^{13/2} \propto P_0^{13/3}$,
so that a planet with a 2 day period is accreted 20 times faster than
a planet with a 4 day period. For a given age of the stellar
population, one would expect to see far more planets in 4 day orbits
than in 2 day orbits (if no observational biases are present).

When stellar damping is included, some of the orbital energy is
dissipated in the planet, and some is dissipated in the star.  For
purposes of finding the impact of energy dissipation on the planet
(for example, the effect on the planetary radius), we need to find the
fraction of the energy that is dissipated in the planet.  For this
case, we can consider both the energy dissipation and the evolution of
semi-major axis to be functions of the eccentricity.  Equation
(\ref{eq:gofe}) specifies the function $g(e)$. The energy dissipated
in the planet, as a function of eccentricity $e$, can be written in
the form 
\be 
\Delta E_P = {G M_\ast m_P \over a_0 (1 - e_0^2) (2 - \Gamma)} 
\Bigl[ e_0^2 - e^2 \bigl( e_0/e \bigr)^\Gamma \Bigr] \, . 
\label{eq:energydiss} 
\ee
This function of eccentricity can be converted into a function of time
using the time evolution equations (\ref{eq:timeint} --
\ref{eq:tevaluate}) discussed above. The self-gravitational energy of
a Jovian planet can be written $E_{GP} = \eta G m_P^2 / R_P$, where
the dimensionless parameter $\eta \approx 5/3$ (although the exact
value depends on the internal structure and of the planet).  We can
thus determine the conditions for which a planet will experience a 
dissipational energy $\Delta E_P$ that is comparable to its 
self-gravity $E_{GP}$: 
\be
{\Delta E_P \over E_{GP} } = {M_\ast \over m_P} {R_P \over  a_0} 
{\Bigl[ e_0^2 - e^2 \bigl( e_0/e \bigr)^\Gamma \Bigr] \over \eta 
(1 - e_0^2) (2 - \Gamma) } \, \approx 2.80 
\Bigl( {e_0^2 \over 1 - e_0^2} \Bigr)
\Bigl( {a_0 \over 0.05 \, {\rm AU} } \Bigr)^{-1} \, . 
\label{eq:energyratio} 
\ee 
In the second approximate equality, we have assumed $M_\ast = 1.0
M_\odot$ and Jovian properties for the planet ($m_P = 1.0 m_J$ and
$R_P = 1.0 R_J$). If a hot Jupiter starts with $a_0$ = 0.05 AU and
eccentricity $e_0 \approx 0.51$, then the amount of energy dissipated
within the planet through tidal interactions is comparable to its
self-gravity, so that substantial structural changes can be forced
upon the planet. In most cases, however, we expect a smaller
eccentricity, so that only a fraction of the planet's
self-gravitational energy would be dissipated (as given by
eq. [\ref{eq:energyratio}]).  For the median of observed eccentricity
of the current sample, $e_M \approx 0.28$, the ratio $\Delta
E_P/E_{GP} \approx$ 0.24, which is still large enough to be
significant.

\subsection{Two planet system with stellar damping term} 

For a two planet system with secular interactions and stellar damping
effects, the equation of motion takes the form
\be
{df \over d\tz} = - {A f^{5/2} \over (\Pi_0 + f^3 - B f^{7/2})^2 
+ C f^{17/2} } \, - \Gamma f^{-11/2} \, . 
\label{eq:tdevolve} 
\ee
If we make the same approximations as before (\S 2.2), where 
$B,C \ll 1$, the equation of motion can be written in terms 
of a formal implicit solution of the form 
\be 
\tz = \int_f^1 \, df \, f^{11/2} \, 
{ (\Pi_0 + f^3)^2 \over \Gamma (\Pi_0 + f^3)^2 + A f^8 } 
\, \, .  
\ee
Although the integral cannot be evaluated in terms of elementary
functions, a good working approximation can be found if we assume that
the first, secular term in the equation of motion is important only
for $f \approx 1$. To find this approximation, we first rewrite the
integral in the form
\be 
\Gamma \tz = \int_f^1 \, df \, \Biggl[ f^{11/2} \, - \,  
{ (A/\Gamma) f^{27/2} \over (\Pi_0 + f^3)^2 + (A/\Gamma) f^8 } 
\Biggr]  \, \, .  
\ee
We then assume that the second term (which arises from secular
perturbations) only is important at the beginning of the evolution
with $f \sim 1$, so we can set $f=1$ in the denominator. The resulting
expression for the evolution time becomes
\be 
\Gamma \tz \approx {2 \over 13} (1 - f^{13/2}) - 
{A / \Gamma \over (\Pi_0 + 1)^2 + A/\Gamma} 
{2 \over 29} (1 - f^{29/2}) \, . 
\label{eq:tdsol} 
\ee 
This approximation is valid as long as the ratio $A/\Gamma$ is not too
large. For example, the error is less than 15\% as long as $A/\Gamma <
4$ for $\Pi_0 = 1$.  In the extreme limit $\Gamma \ll 1$, where this
approximation fails, the additional damping term does not play a role
and we can use the no damping solution as a good working
approximation.  Comparison of this result with that of the previous
subsection indicates that in order for the secular interaction of the
second planet to play a role, the second term must compete with the
first. Almost equivalently, the first term in the evolution equation
(\ref{eq:tdevolve}) must compete with the second, i.e., $A \sim \Gamma
(1 + \Pi_0)^2$. Inserting typical masses and radii for the planets and
star, the requirement for secular interactions to dominate the long
term evolution becomes
\be 
15 \, e_2 \Bigl( {a_1 \over a_2} \Bigr) \, 
\Bigl( {Q_\ast \over Q_P} \Bigr)^{1/2} \gta 
1 + (7.9 \times 10^{-4}) \Bigl( {a_1 \over 0.05 {\rm AU}} \Bigr) \, 
\Bigl( {a_1 \over a_2} \Bigr)^{-3} \, . 
\ee 
Since $a_1 \ll a_2$ for typical systems, this requirement is not 
met unless $Q_\ast \gg Q_P$. As a result, the stellar damping effects 
tend to dominate secular interactions as the inner planet migrates 
closer to the central star. 

For solar systems with sufficiently small semi-major axes and high
eccentricities, secular interactions coupled with stellar damping
drive the inner planet into the star. To define a region of parameter
space for which this effect is important, we take the inner planet to
be a hot Jupiter with $m_1 = m_J$, $a$ = 0.05 AU, and an eccentricity
that cycles through $e_1$ = 0.  The orbital elements for the second
planet that lead to planetary accretion within 5 Gyr is shown in
Figure \ref{fig:longterm2}, where we have taken $Q_\ast$ = $10^6$.
Since the effect depends on the mass of the second planet, we show
curves for $m_2$ = 1, 3, and 5 $m_J$.  This plot is thus the analog of
that shown in Figure \ref{fig:longterm1}, which does not include the
stellar damping term. As expected, inclusion of stellar damping
effects leads to a larger region of parameter space for which planets
can be accreted. The dashed curve delineates the portion of parameter
space for which a system with $m_2 = m_J$ would drive its inner planet 
into the star within 5 Gyr in the absence of relativity. 
 
\section{Conclusion}  

This paper presents solutions for the the long term evolution of four
types of planetary systems --- one and two planet systems with tidal
circularization, both with and without the inclusion of stellar
damping effects. For the two planet systems, planet-planet
interactions are modelled using secular interaction theory (MD99)
including leading order relativistic corrections (Paper I).  The
solutions are presented in implicit form; for example, $\tz (f)$ gives
the time as a function of the inner planet's semi-major axis, as
specified by $f = a_1(t)/a_1(0)$. In the absence of stellar damping,
we find exact analytic solutions to the long term evolution, as given
by equations (\ref{eq:jsol}) and (\ref{eq:twosol}). For systems in
which the stellar damping term is important, we find approximate
analytic solutions given by equations (\ref{eq:timesol}) and
(\ref{eq:tdsol}).

\begin{figure} 
\figurenum{1} 
\centerline{ \epsscale{0.8} \plotone{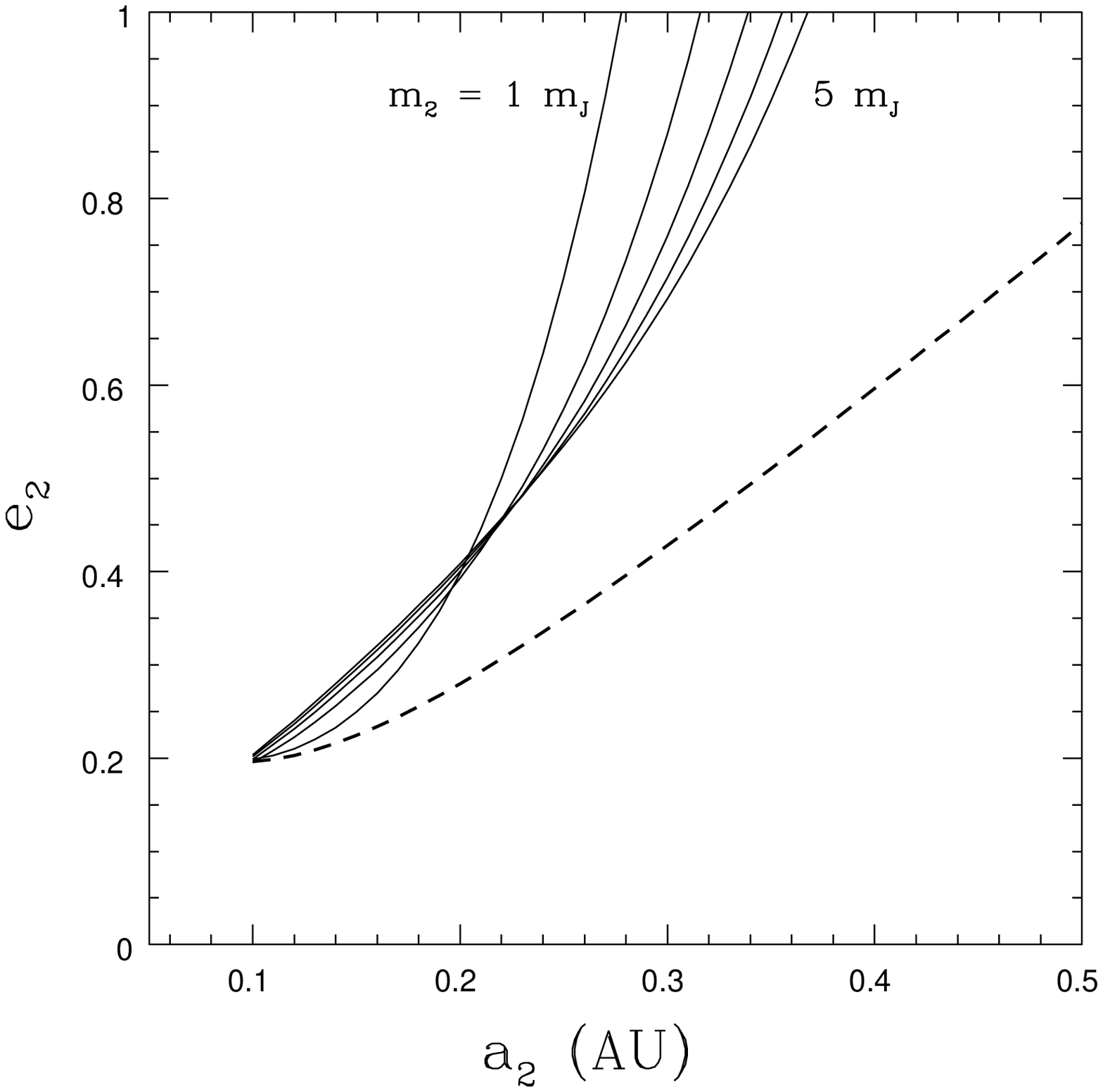} } 
\caption{ For systems that contain a hot Jupiter, this plot
shows the orbital elements of the second planet $(a_2,e_2)$ required
to drive the hot Jupiter into the stellar photosphere over a time
scale of 5 Gyr. All cases assume that the inner planet has mass $m_1 =
m_J$, semi-major axis $a_1$ = 0.05 AU, and an eccentricity that cycles
through $e_1$ = 0 (where $\ebar \ne 0$ due to secular interactions
with the second planet). The five solid curves correspond to five
choices of the mass of the second planet, as labeled. The region of
parameter space above the curves corresponds to systems in which the 
inner planet is driven into the star on shorter time scales $t < 5$
Gyr. The dashed curve delineates the (much larger) region of parameter 
space for which the inner planet would be driven into the star in the 
absence of general relativity (for $m_2 = m_J$). }
\label{fig:longterm1} 
\end{figure}

\begin{figure} 
\figurenum{2} 
\centerline{ \epsscale{0.8} \plotone{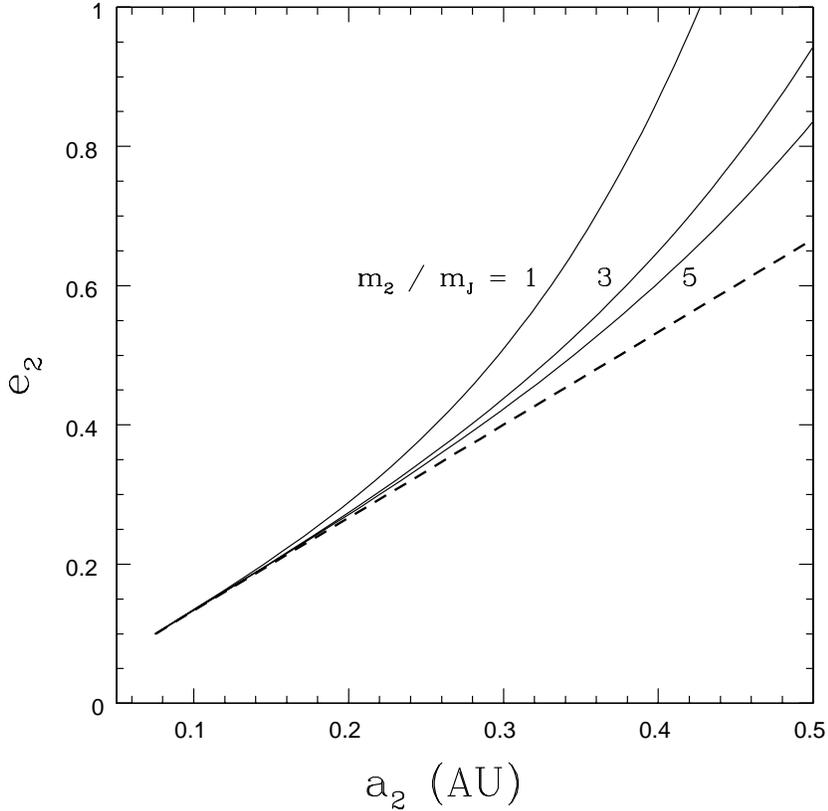} } 
\caption{ For systems that contain a hot Jupiter, this plot
shows the orbital elements of the second planet $(a_2,e_2)$ required
to drive the inner planet into the star within 5 Gyr, where a stellar
damping term is included with $Q_\ast$ = $10^6$ (compare with Fig.
\ref{fig:longterm1}).  The inner planet has mass $m_1 = m_J$,
semi-major axis $a_1$ = 0.05 AU, and an eccentricity that cycles
through $e_1$ = 0 (where $\ebar \ne 0$ due to secular interactions
with the second planet). The three solid curves correspond to
different masses of the second planet, as labeled. The region of
parameter space above the curves corresponds to systems in which the 
inner planet is driven into the star on time scales $t < 5$ Gyr. 
The dashed curve delineates the region of parameter space for 
which the inner planet would be driven into the star in the absence  
of general relativity (for $m_2 = m_J$). }
\label{fig:longterm2} 
\end{figure}

In multiple planet systems, secular interactions enhance the accretion
of inner planets by the central star, provided that the outer planet
has a sufficiently large mass, small semi-major axis, and/or large
eccentricity. We have presented a quantitative assessment of the
orbital elements of a second planet required to drive a hot Jupiter
into the central star within 5 Gyr, for two-planet systems both with
(Fig. \ref{fig:longterm2}) and without (Fig. \ref{fig:longterm1})
stellar damping. General relativity acts to delay the accretion of
planets, i.e., the region of parameter space for which the inner
planet would be accreted within a given time would be much larger in
the absence of relativistic effects (see the dashed curves in Figs. 
\ref{fig:longterm1} and \ref{fig:longterm2}). 

For multiple planet systems with hot Jupiters as inner planets,
secular interactions tend to give the inner planet nonzero
eccentricity. This continual addition of eccentricity, in conjunction
with the circularization processes experienced by such planets, leads
to large amounts of energy dissipation within the hot Jupiters (eqs.
[\ref{eq:energydiss}] and [\ref{eq:energyratio}]).  Such extreme
dissipation, in turn, may lead to mass loss and thereby explain the
relatively small masses observed for close planetary companions. This
process should be studied in greater detail in the future.

The framework of analysis presented here can be readily applied to
individual systems whenever the orbital elements of the constituent
planets are known to reasonable accuracy. We maintain an up-to-date
catalog of the known extrasolar
planets\footnote{http://www.ucolick.org/\~{$\,$}laugh/} where the time
scales derived here are tabulated using the best available fits to the
radial velocity data sets. This data base, in conjunction with the
analytic solutions presented here, should provide a useful resource 
for further research on close planetary systems. 

\bigskip 

We would like to thank an anonymous referee for suggesting that this
paper be published separately from Paper I. This work was supported at
U. Michigan (FCA) by the Michigan Center for Theoretical Physics and
by NASA through the Terrestrial Planet Finder Mission (NNG04G190G) and
the Astrophysics Theory Program (NNG04GK56G0).  This material is based
in part upon work supported by the National Science Foundation CAREER
program under Grant No. 0449986 (GL), and was also supported at
U. C. Santa Cruz (GL) by NASA through the Terrestrial Planet Finder
Precursor Science Program (NNG04G191G) and through the Origins of
Solar Systems Program (NAG5-13285).

\newpage 

\end{document}